\documentclass{ifacconf}
\usepackage{amsmath,amssymb,amsfonts}
\usepackage{graphicx}      
\usepackage{natbib}        
\usepackage{xcolor}
\usepackage{tabu}
\usepackage{booktabs}
\usepackage{subcaption}
\usepackage{tikz}
\usepackage{pgfplots}
\usepackage{url}

\pgfplotsset{
    compat=1.18
}
\begin{document}
\begin{frontmatter}

\title{\vspace{-10pt} Clustering Techniques for Stable Linear Dynamical Systems with applications to Hard Disk Drives} 


\author[First]{Nikhil Potu Surya Prakash} 
\author[First]{Joohwan Seo} 
\author[Second]{Jongeun Choi}
\author[First]{Roberto Horowitz}

\address[First]{University of California, Berkeley, CA, 94720 USA}
\address[Second]{Yonsei University, Seoul, Republic of Korea
\\ e-mail: \{joohwan\_seo, nikhilps, horowitz\}@berkeley.edu, 
jongeunchoi@yonsei.ac.kr
}

\begin{abstract} 
In Robust Control and Data Driven Robust Control design methodologies, multiple plant transfer functions or a family of transfer functions are considered and a common controller is designed such that all the plants that fall into this family are stabilized. Though the plants are stabilized, the controller might be sub-optimal for each of the plants when the variations in the plants are large. This paper presents a way of clustering stable linear dynamical systems for the design of robust controllers within each of the clusters such that the controllers are optimal for each of the clusters. First a k-medoids algorithm for hard clustering will be presented for stable Linear Time Invariant (LTI) systems and then a Gaussian Mixture Models (GMM) clustering for a special class of LTI systems, common for Hard Disk Drive plants, will be presented. 

\end{abstract}


\end{frontmatter}

\section{Introduction}\label{sec:Introduction}
Transfer function clustering is a useful technique in control systems engineering because it allows us to identify groups of transfer functions that have similar characteristics, which can help us better understand the behavior of a system and design appropriate control strategies. 

Some sub-fields of dynamical systems where we might want to cluster transfer functions are as follows: 
\begin{enumerate}

\item System identification: Clustering transfer functions can help us identify different modes of a system, which can aid in model identification and system identification. By clustering transfer functions with similar behavior, we can identify distinct system modes and their associated dynamics.

\item Control design: Clustering transfer functions can help us design control strategies that are tailored to specific modes of a system. For example, if we have a system with multiple modes, we might design a control strategy that switches between different controllers depending on the current mode.



\end{enumerate}

Overall, clustering transfer functions can provide valuable insights into the behavior of a system, and can help us design effective control strategies and diagnose faults. In this paper, the focus will be on clustering stable linear dynamical systems common for hard disk drive (HDD) plants. Once the clusters have been finalized, controllers can be designed for each of the plants with existing control design techniques so that the controllers are both optimal and robust within each cluster. Though the focus will be on stable LTI systems, an idea on how this technique can be extended to unstable LTI systems will be presented.

Recently, in most of the personal computers, Solid State Drives (SSD) have replaced HDDs as the data transfer rate of SSDs is much higher than that of HDDs. Despite this, HDDs continue to dominate the market in data centers due to their lower cost and higher reliability when compared to SSDs. The cost of memory per square inch of an HDD can be decreased by decreasing the spacing between successive tracks on which the data is stored. One of the factors that dictates the spacing between the tracks in a HDD is the sensitivity of the controller to disturbances. To improve the disturbance rejection characteristics of HDDs, additional actuators are added. Voice Coil Motor (VCM) actuator is responsible for compensating low frequency disturbances and has higher strokes while piezo-electric micro actuators (PZT) are responsible for compensating high frequency disturbances and have lower strokes typically around one to three tracks. A dual stage actuator (DSA) system has a VCM and a PZT actuator while a triple stage actuator (TSA) system has a VCM and two PZTs. 


Recently, data-driven feedback control design approaches from frequency response measurements ((\cite{GaKaLo:10,KaKa:17,KaNiZh:16})) for multiple stage systems have been developed in \cite{Ba:17,BaShHo:18, Ni:22, Ni:23} to suppress the disturbances. These data-driven controllers are robust to the variations in the plant models. Though robustness is ensured when a solution to the optimization problem exits, a common controller might not be the optimal controller for each individual HDD as there might be variations in the plants and the disturbances. To address these issues, data driven feedforward controllers have been developed in \cite{ShHo:20,ShHo:19,ShHo:21} based on the frequency response measurement of disturbance processes on top of the robust feedforward controllers and another add-on adaptive feedforward  controller has been developed in \cite{ChPrHo:22} on top of the robust feedback controller to account for plant variations. Robust controllers for Linear Time Invariant (LTI) plants with parametric uncertainties with applications to HDDs have also been developed in \cite{Ch:06,Ch:08} utilizing plant models instead of a data driven approach.

Performance can be improved while still keeping the controllers robust to plant variations by clustering the plants into different clusters and designing a controller for each cluster separately such that the controller is optimal within the cluster. Since every cluster contains other plants, robustness is also imposed. Once the controllers are designed, controllers for each system that comes out the manufacturing plant can be selected by using time responses/frequency responses of the system to decide which cluster the system belongs to.


The flow of this paper is organized as follows: In section~\ref{sec:PreviousWork}, we present some of the distance metric used for clustering in the literature and list their shortcomings. In section~\ref{sec:Preliminaries}, we present some of the preliminaries for $\mathcal{H}$ norms of systems and how they can be used to define distances between two linear dynamical systems. In section~\ref{sec:HC}, we present the k-medoids algorithm for hard clustering LTI systems and in section~\ref{sec:SC}, we present a soft clustering algorithm utilizing Gaussian Mixtures Models clustering for a class of plans which are common for Hard Disk Drives. Finally in section~\ref{sec:results}, we show the effectiveness of the algorithm in clustering HDD plants.

\section{Previous Work}\label{sec:PreviousWork}
The most critical part of clusterings, such as k-means or k-medoids, is the notion of a distance between two data points. Each data point in this paper represents an LTI system. It could either be a transfer function or a set of frequency responses. One of the first notions of distances between two linear dynamical systems was presented in \cite{Ma:00} with the distance defined between two ARMA models. Let $H_1(z^{-1})$ and $H_2(z^{-1})$ be the z-transforms of two discrete time stable LTI systems of the same order.  
One of the major drawbacks of this distance metric is that in certain cases, two LTI dynamical systems which produce the same output due to pole-zero cancellations might have a non-zero distance between them according to the above distance metric. 
\cite{Ci:14} proposed a distance metric in terms of the system matrices. If the realizations of strictly proper stable LTI systems $H_1$ and $H_2$ are $(A_1,B_1,C_1)$ and $(A_2,B_2,C_2)$ respectively, then, the distance $d(H_1,H_2)$ between them with positive weights $\lambda_A$, $\lambda_B$ and $\lambda_C$ is defined as
\begin{equation}
    d(H_1,H_2)^2 = ||C_1-C_2||^2_F+\lambda_A||A_1-A_2||^2_F+\lambda_B||B_1-B_2||^2_F
\end{equation}
Such a distance is defined only when both systems have the same order. Since multiple realizations are possible for the same system, such a distance metric would produce non zero distances between two different realizations of the same system. \newline
To overcome the shortcomings of these distance metrics, we propose a distance metric in terms of the $\mathcal{H}$ norms of the system, the details of which are presented in the upcoming sections.

\section{Preliminaries}\label{sec:Preliminaries}
In this section, a few mathematical preliminaries necessary for clustering stable LTI dynamical systems will be provided.
\subsection{System Norms}\label{subsec:SN}
For a continuous time transfer function $G_{m \times n}(s)$ or a discrete time transfer function $G_{m \times n}(z)$ with sampling time $T_s$ with $m$ inputs and $n$ outputs, two types of system norms are defined namely $H_\infty$ norm and $H_2$ norm. Let the realization of a continuous time system $G$ in terms of system matrices $A,B,C$, and $D$ be denoted as 
\begin{equation}
    G \sim\ \left[
\begin{array}{c|c}
A & B \\
\hline
C & D
\end{array}
\right]
\end{equation}
\subsubsection{$H_\infty$ Norm: }\label{subsec:hinfinity}
The $H_\infty$ norm is defined as the supremum of the largest singular value of the transfer function across the entire frequency range. \newline
For continuous time systems, the $H_\infty$ norm is defined as 
\begin{equation}\label{eq:hinftynormct1}
    ||G_{m \times n}||_\infty = \underset{\omega \in \Omega}{\sup} \; \Bar{\sigma}(G_{m \times n}(j\omega))
\end{equation}
where $\Omega = (-\infty,\infty)$ is the entire frequency range and $\Bar{\sigma}(\cdot)$ is the largest singular value. \newline
By utilizing the definition of matrix norms, the $H_\infty$ norm can also be equivalently written as
\begin{equation}\label{eq:hinftynormct2}
    ||G_{m \times n}||_\infty = \underset{\omega \in \Omega}{\sup} \; ||G_{m \times n}(j\omega)||_2
\end{equation}
For Single Input Single Output (SISO) systems, the definition boils down to 
\begin{equation}\label{eq:hinftynormctsiso}
    ||G_{1 \times 1}||_\infty = \underset{\omega \in \Omega}{\sup} \; |G_{m \times n}(j\omega)|
\end{equation}

Similarly, for discrete time systems, the $H_\infty$ norm is defined as
\begin{equation}\label{eq:hinftynormdt1}
    ||G_{m \times n}||_\infty = \underset{\omega \in \Omega}{\sup} \; \Bar{\sigma}(G_{m \times n}(e^{j\omega T_s})) = \underset{\omega \in \Omega}{\sup} \; ||G_{m \times n}(e^{j\omega T_s})||_2
\end{equation}
where $\Omega = (-\frac{\pi}{T_s},\frac{\pi}{T_s}]$ denotes the entire frequency range upto the Nyquist frequency.
For SISO systems, the definition boils down to
\begin{equation}\label{eq:hinftynormdtsiso}
    ||G_{1 \times 1}||_\infty = \underset{\omega \in \Omega}{\sup} \; |G_{m \times n}(e^{j\omega T_s})|
\end{equation}
For SISO systems, it can be seen that the $H_\infty$ norm is the peak of the bode magnitude plot which is also the same as the maximum amplification of an input sinusoidal wave for stable systems. Therefore
\begin{equation}\label{eq:hinftynormtd}
    ||G_{m \times n}||_\infty = \underset{u \in \mathcal{L}_2, ||u||_2\geq 1}{\sup} \; \frac{||y||_2}{||u||_2}
\end{equation}
where the space $\mathcal{L}_2$ is the space of all square integrable functions.\\
Instead of sweeping the entire frequency range, the $H_\infty$ norm can be obtained from the state matrices by utilizing the bounded real lemma. For this, a Hamiltonian matrix ($\mathcal{H}$) is constructed as follows (\cite{Ku:13}).
\begin{equation}
\mathcal{H} = \left[ \begin{array}{cc}
         A & \frac{BB^T}{\gamma^2} \\
         -C^TC & -A^T 
    \end{array} \right]
\end{equation}
The smallest $\gamma$ that keeps the Hamiltonian matrix asymptotically stable is the $H_\infty$ norm of the system. This value is usually found using an iterative scheme like the bisection technique. An equivalent result exists for discrete time systems as well.
\subsubsection{$H_2$ Norm:}\label{subsubsec:h2}
$H_2$ norm of a system can be viewed as the energy of the dynamical system. For SISO systems, it is a representation of the area under the bode plot. The $H_2$ norm of a system is finite only when the system is stable and the feedthrough matrix $D$ is zero i.e., the system must be strictly proper. The $H_2$ norm for a continuous time  Multiple Input Multiple Output (MIMO) system is defined as
\begin{subequations}\label{eq:h2norm1}
\begin{align}
    ||G_{m \times n}||^2_2 &= \frac{1}{2 \pi} \int_{-\infty}^{\infty}tr[G(j\omega)^*G(j\omega)]d\omega \\
    &= \frac{1}{2 \pi} \int_{-\infty}^{\infty}tr[G(-j\omega)^TG(j\omega)]d\omega
\end{align}
\end{subequations}
where $(\cdot)^*$ represents the complex conjugate. For systems whose frequency responses are available instead of the model, the approximate $H_2$ norm is given as
\begin{equation}
   ||G_{m \times n}||^2_2 \approx \frac{1}{2\pi}\sum_{k=1}^{N}tr[G(j\omega_{k})^{*}G(j\omega_k)]\Delta \omega_{k}
\end{equation}
where $\omega_k$ is the $k^{th}$ frequency at which the response is available.\newline
Utilizing Parseval's theorem, the $H_2$ norm in \eqref{eq:h2norm1} can be written in terms of the system matrices as
\begin{equation}\label{eq:trCPCt}
    ||G_{m \times n}||^2_2 = tr[CPC^T]
\end{equation}
where $tr[\cdot]$ denotes the trace and the matrix $P$ is the controllability grammian given by 
\begin{equation}\label{eq:P}
    P = \int_{-\infty}^{\infty}e^{At}BB^Te^{A^Tt}dt
\end{equation}
which is also the solution of the Lyapunov equation given by
\begin{equation}
    AP+PA^T = -BB^T
\end{equation}
We can also express \eqref{eq:trCPCt} in terms of the input matrix $B$, and in that case, the grammian would be the observability grammian. $H_2$ norms can also be calculated for discrete time systems in a similar fashion.
\subsection{$H_2 \mbox{ and } H_\infty$ Distances}
These norms can now be used to define the notion of another distance between transfer functions which can then be used to cluster transfer functions. Let us consider two stable continuous time MIMO systems $G_1$ and $G_2$ with state space realizations as follows 
\begin{equation}
G_1 \sim\
\left[
\begin{array}{c|c}
A_1 & B_1 \\
\hline
C_1 & D_1
\end{array}
\right] \; , 
G_2 \sim\
\left[
\begin{array}{c|c}
A_2 & B_2 \\
\hline
C_2 & D_2
\end{array}
\right], 
\end{equation}
The difference between these two transfer functions has the following state space realization
\begin{subequations}
\begin{align}
G_1 - G_2 & \sim\
\left[
\begin{array}{c c|c}
A_1  &0 & B_1 \\
0    &A_2 & B_2 \\
\hline
C_1 \; &-C_2 & D_1-D_2
\end{array}
\right] \\
&= 
\left[
\begin{array}{c|c}
A_{12}   & B_{12} \\
\hline
C_{12}  & D_{12}
\end{array}
\right]
\end{align}
\end{subequations}
The idea behind using the difference is to see how much the outputs of the systems differ under the same input. \newline
If $D_{12} = 0$, the $H_2$ distance between the transfer functions $G_1$ and $G_2$ exists and can be found as follows 
\begin{equation}
    ||G_1-G_2||^2_2 = tr[C_{12}P_{12}C_{12}^T] 
\end{equation}
where 
\begin{equation}
    P_{12} = \int_{-\infty}^{\infty}e^{A_{12}t}B_{12}B_{12}^Te^{A_{12}^Tt}dt
\end{equation}
which is the solution of the Lyapunov equation given by
\begin{equation}
    A_{12}P_{12}+P_{12}A_{12}^T = -B_{12}B_{12}^T
\end{equation}

The $H_\infty$ distance between the transfer functions $G_1$ and $G_2$ can be found by constructing the Hamiltonian matrix as
\begin{equation}
\mathcal{H}_{12} = \left[ \begin{array}{cc}
         A_{12} & \frac{B_{12}B_{12}^T}{\gamma^2} \\
         -C_{12}^TC_{12} & -A_{12}^T 
    \end{array} \right]
\end{equation}
The smallest $\gamma$ that keeps the Hamiltonian matrix $\mathcal{H}_{12}$ asymptotically stable is the $H_\infty$ distance between the transfer functions $G_1$ and $G_2$.

For systems whose frequency responses are available instead of the model, the approximate $H_2$ distance is given as
\begin{equation}
   ||G_1\text{-}G_2||^2_2 \approx \frac{1}{2\pi}\sum_{k=1}^{N}tr[(G_1\text{-}G_2)(j\omega_{k}))^{*}((G_1\text{-}G_2)(j\omega_{k})]\Delta \omega_{k}
\end{equation}
In addition, the approximate $H_\infty$ distance can be calculated as
\begin{equation}\label{eq:hinftynormct2}
    ||G_1-G_2||_\infty = \underset{\omega_k \in \{\omega_1, \dots , \omega_N\}}{\sup} \; ||G_1(j\omega_k)-G_2(j \omega_k)||_2
\end{equation}

Though these norms and distances are defined for stable LTI systems, we can extend the notion of the distance to unstable LTI systems by first designing a common controller for all the systems and using the same notions on the closed loop systems. Note that this distance would not be representative of how much the outputs differ as the outputs diverge for unstable systems in open loop. Nevertheless, when a stabilizing controller exists, we can use the distances for the closed loop systems to cluster the open loop plants. But it is possible that such a stabilizing controller might not exist for all plants that we are interested in clustering.

\section{Hard Clustering}\label{sec:HC}
In this section, utilizing the distance metrics, the transfer functions/ frequency responses can be clustered using a k-medoids algorithm with the chosen distance metric. We prefer a k-medoids algorithm to a k-means algorithm for two reasons:
\begin{itemize}
    \item Robustness to outliers - k-medoids is more robust to outliers than k-means because medoids are selected as the most centrally located point in a cluster, which is less sensitive to outliers than the mean. This means that k-medoids can handle noisy data better than k-means.
    \item Defining a mean frequency response is easy for data as a simple arithmetic mean serves as the center of mass. But for transfer functions, the mean transfer function obtained by summing all the given transfer functions and dividing by the number of data points would produce a very high dimensional transfer function. Computing the distances from this mean to other data points might become computationally very expensive. However, it is still theoretically possible to have such a mean.
\end{itemize}
 The mean system obtained by taking the average of $N$ systems with $(A_i,B_i,C_i,D_i)$ representing the system matrices of the $i\textsuperscript{th}$ system is   
\begin{equation}\label{eq:meansys}
\frac{1}{N} \sum_{i=1}^{N} G_i \sim \left[
\begin{array}{c c c c|c}
A_1 & 0 & \ldots & 0 & B_1 \\
0 & A_2 & \ldots & 0 & B_2 \\
\vdots & \vdots & \ddots & \vdots & \vdots \\
0 & 0 & \ldots & A_N & B_N \\
\hline
\frac{C_1}{N} & \frac{C_2}{N} & \ldots & \frac{C_N}{N} & \frac{1}{N}\sum_{i=1}^{N} D_i
\end{array}
\right]
\end{equation}    
It can easily be seen that the order of the mean system scales linearly with the number of data points. 

Therefore, to improve robustness and more importantly to keep the computational effort reasonable, instead of finding the mean, we find the point in the group that minimizes the total sum of distances to the other points.

\section{Soft Clustering}\label{sec:SC}
Soft clustering allows for a data point to belong to multiple clusters to varying degrees, while, in contrast, hard clustering assigns a data point to only one cluster. In soft clustering, each data point is assigned a probability of belonging to each cluster, with the sum of probabilities across all clusters equaling one. These probabilities are based on the distance between the data point and the cluster centers, and the degree of overlap between the clusters. These are useful in situations where there is ambiguity or uncertainty in the data, or when a data point may belong to multiple groups simultaneously. The advantage of soft clustering is that it can be used to design probabilistic controllers. In this section, we utilize a Gaussian Mixtures Model (GMM) soft clustering on a feature vector that summarizes the plant. A feature vector is used as an alias for the entire plant and a covariance matrix can be defined for the feature vectors whereas a notion of covariance does not make sense for the frequency response data. Instead of GMMs, one could also use a technique like fuzzy C-means clustering (\cite{La:14}) directly on the data points, without the intermediate step of finding the feature vector, which just requires the notion of a distance. A fuzzy C-means algorithm can be used with the $\mathcal{H}$ distances defined in section \ref{sec:Preliminaries}.
\begin{figure}[h!]
    \centering
    \includegraphics[width = 0.5\textwidth]{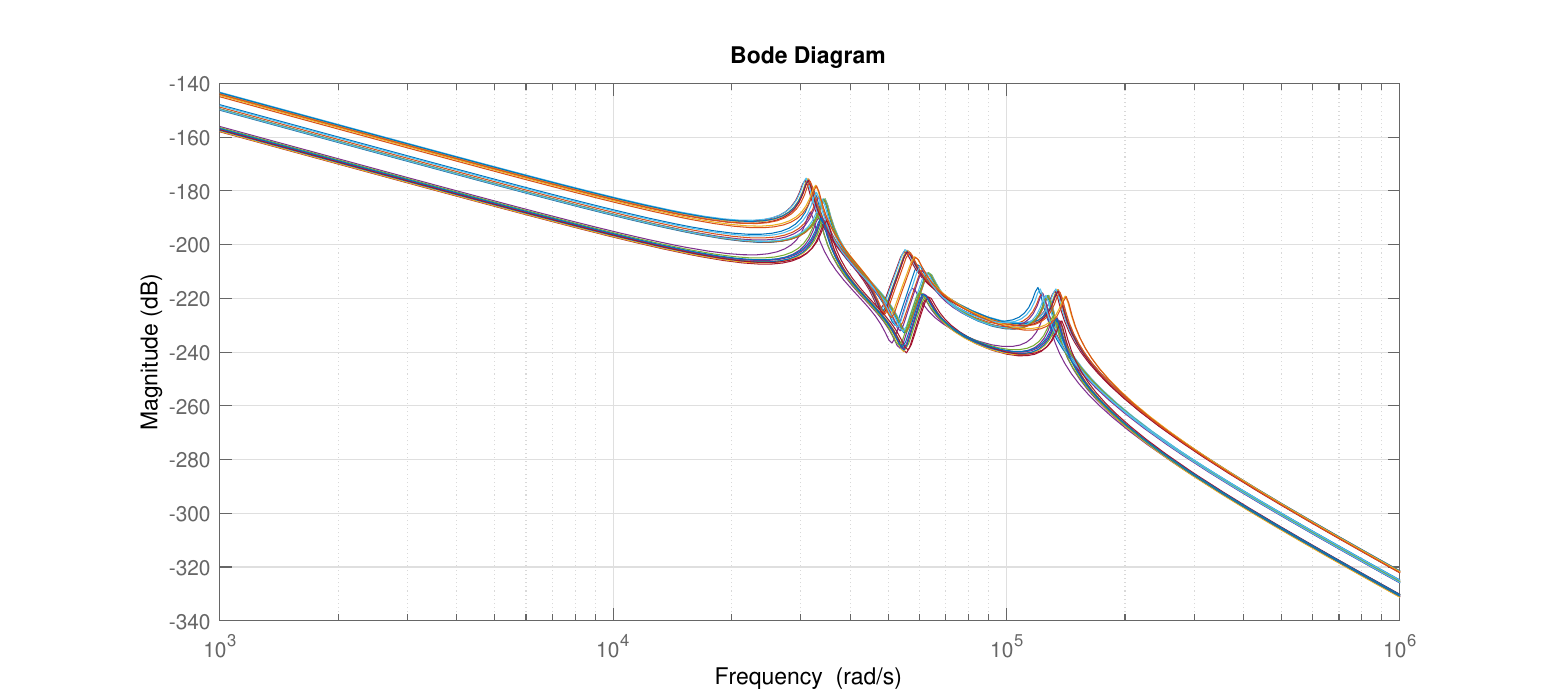}
    \caption{Frequency responses of 30 VCM plants}
    \label{fig:allplants}
\end{figure}
\begin{figure}[h!]
    \centering
    \includegraphics[width = 0.5\textwidth]{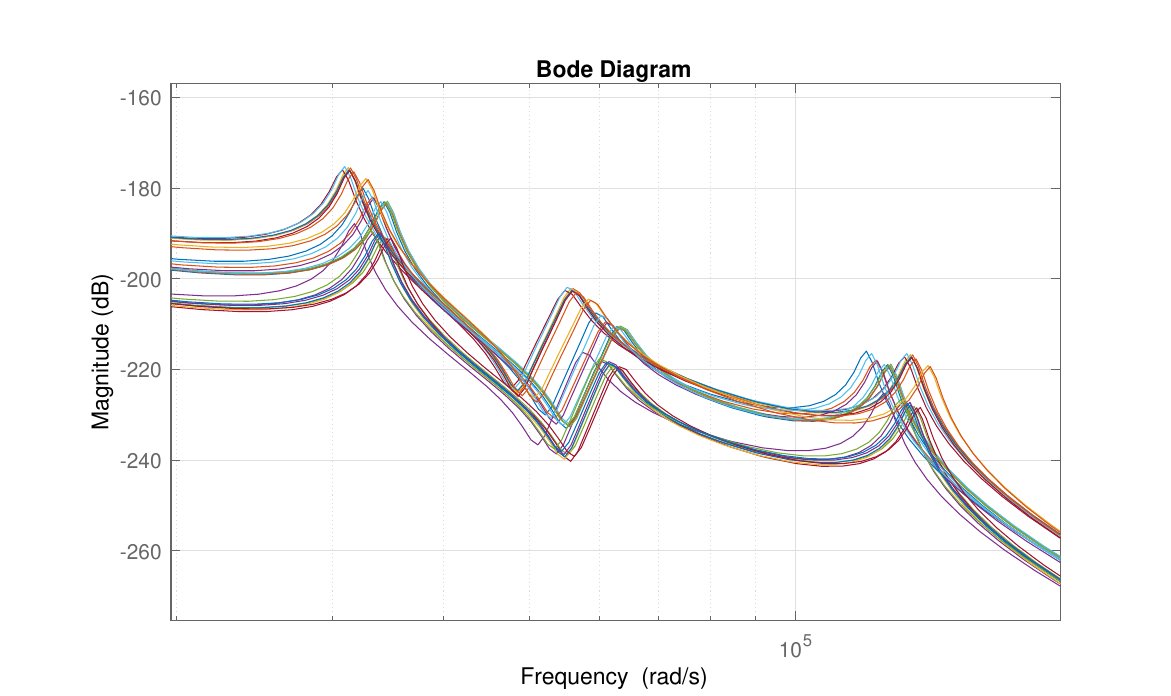}
    \caption{Zoomed frequency responses of the 30 VCM plants in Fig.\ref{fig:allplants} around the resonant peaks.}
    \label{fig:allplantszoomed}
\end{figure}

The frequency response data for a batch of 30 VCM plants, is depicted in Fig. \ref{fig:allplants}. The data shows several discernible resonance modes. The data has been generated from HDD plant models representative of real plants. The various modes in the increasing order of natural frequency are Butterfly mode at around 1.9 KHz, Torsion mode around 2.3 KHz and Sway mode around 16.5 KHz. Hence, it is suitable to represent the transfer function between the input and output as a sum of modes, which can be expressed as follows:
\begin{equation}
G(s) = \frac{b_0}{s^2}+\sum_{k=1}^{n}\frac{b_k}{s^2+2\zeta_k\omega_ks+w_k^2}
\end{equation}
In this section, we stick to plants of this form, but soft clustering can also be done for other stable LTI plants using fuzzy C-means clustering algorithm. \newline
To represent the transfer function as a sum of modes, only three parameters per mode are needed: natural frequency, damping coefficient and modal constant. The parameters for each mode in the summation form can be easily determined through a single degree of freedom (SDOF) modal analysis. The SDOF model is based on the assumption that the dynamics of a system near a resonance frequency are mainly dominated by the corresponding resonance mode, and other modes contribute to a lesser extent. This assumption is generally valid for disk drive structures (\cite{Co:07}).

\begin{figure}[h!]
    \centering
    \includegraphics[width = 0.5\textwidth]{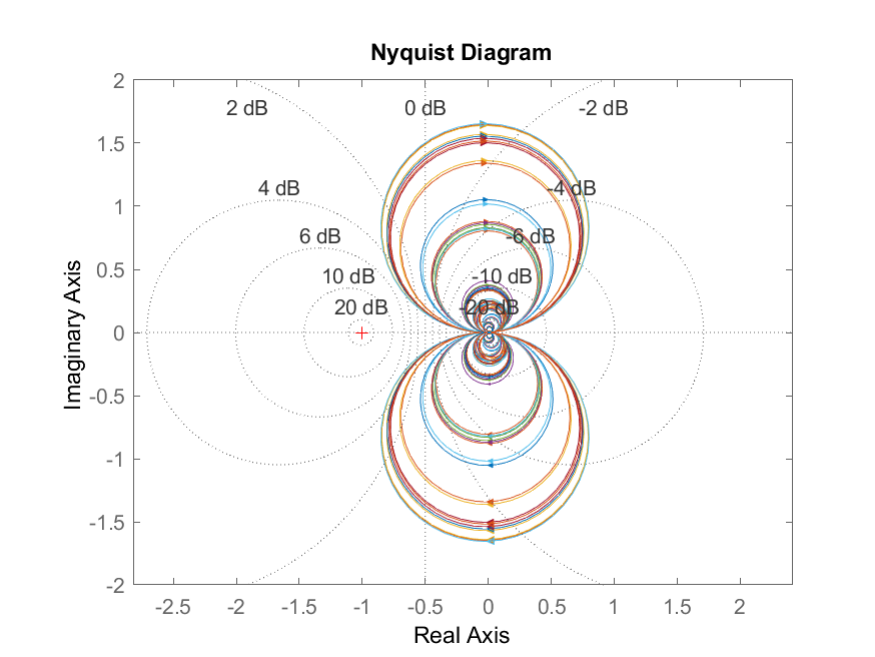}
    \caption{Nyquist plots of the 30 VCM plants in fig.\ref{fig:allplants} showing the circles in the complex plane near the resonant frequencies.}
    \label{fig:nyquistallplants}
\end{figure}

The circular shape of the data points in the Nyquist plane in fig. \ref{fig:nyquistallplants} is utilized to identify the parameters, i.e., natural frequency, damping coefficient and modal constant, as these can be extracted from the circle's geometry. A least-square algorithm can be used to fit the circle, making the process easier. We will just present the equations that are used to find the modal parameters from the frequency responses of the plant and the reader is referred to \cite{Co:07, Ew:86} for more details about the method.

The frequency at which the maximum sweep rate of data points around the circle is attained is known as the natural frequency, and it is defined by the expression:
\begin{equation}
    \gamma(\omega^2) = \frac{d\theta}{\d(\omega^2)} \approx \frac{\Delta \theta}{\Delta (\omega^2)}
\end{equation}

The other modal parameters can be determined by using the following equations:
\begin{subequations}
\begin{align}
\eta = \frac{2}{\omega_n^2 \gamma_{max}} \\
b_0 = 2R\omega_n^2\eta
\end{align}
\end{subequations}

This process is repeated for all the modes by first picking all the local peaks and using the circle fit method around these local peaks. All the model parameters can be collected to form a feature vector $\phi$ for each plant as follows
\begin{equation}
    \phi = [b_0,b_1 \dots , b_n, \zeta_1, \dots, \zeta_n, \omega_1, \dots, \omega_n]^T
\end{equation}

Now, a GMM can be used to cluster the plants based on these feature vectors.


\subsection{Gaussian Mixtures Model Clustering}
Gaussian Mixtures Model (GMM) clustering is a statistical method for clustering data points into multiple groups or clusters based on their similarity. In this method, each cluster is modeled as a probability distribution with a Gaussian (normal) distribution. A GMM assumes that the data points in each cluster are generated from a mixture of multiple Gaussian distributions with different means and variances. The GMM clustering algorithm estimates the parameters of these Gaussian distributions (such as means and variances) and assigns each data point to the cluster with the highest probability of generating that point. 
\begin{equation}
    P_\Phi(\phi) = \sum_{k=1}^{K} \pi_k P_{\Phi|Z=k}(\phi;\mu_k,\Sigma_k)
\end{equation}
where 
\begin{subequations}
\begin{align}
    &\phi: &&\mbox{is the observed quantity} \nonumber \\
    &Z: &&\mbox{is one of K classes} \nonumber \\
    &\Phi|Z=k: &&\mbox{is Gaussian} \nonumber \\
    &\mu_k: &&\mbox{is the mean of cluster K} \nonumber \\
    &\Sigma_k: &&\mbox{is the covariance of cluster K} \nonumber \\
    &\pi_k: && \mbox{is the probability of belonging to cluster k } \nonumber \\
    &P_{\Phi}: && \mbox{is the probability density of } \phi \nonumber    
\end{align}
\end{subequations}
with
\begin{equation}
    \begin{aligned}
    &P_{\Phi|Z=k}(\phi,\mu_k,\Sigma_k) \\ &\quad = 
    \frac{1}{(2\pi)^{D/2}|\Sigma_k|^{1/2}}\exp(-\frac{1}{2}(\phi-\mu_k)\Sigma_k^{-1}(\phi-\mu_k))
    \end{aligned}
\end{equation}
Maximum Likelihood Estimate (MLE) for the Guassian Mixtures is given by
\begin{equation}\label{eq:theta}
    \theta = (\pi_1,\mu_1,\Sigma_1, \pi_2, \mu_2, \Sigma_2, \dots, \pi_K,\mu_K,\Sigma_K)
\end{equation}
which is the solution of the following optimization problem 
\begin{equation}\label{eq:EM}
\begin{aligned}
& \underset{\theta}{\text{maximize}}
& &  \sum_{i=1}^{N} \ln(\sum_{k=1}^{K}\pi_k P_{\Phi|Z=k}(\phi_i,\mu_k, \Sigma_k)) \\
& \text{subject to}
& & \sum_{k=1}^{K}\pi_k = 1 \\
& & & \pi_k \geq 0 \\
& & & \Sigma_k \succ 0. 
\end{aligned}
\end{equation}
The above optimization problem can be solved by the Expectation Maximization (EM) algorithm (\cite{Ya:12}).

\section{Results}\label{sec:results}
In this section, we present the results of hard and soft clustering algorithms applied to the frequency response data of VCM plants shown in fig. \ref{fig:allplants}. It can be seen that the responses can be categorized into three different clusters. But for the algorithm, we used the elbow method to pick the number of optimal clusters from Hard Clustering. Both the algorithms were successfully able to cluster the plants into three clusters.
For the GMM approach, feature vectors of size ten (one feature corresponding to the rigid body mode, three features corresponding to each resonant mode) were constructed from the frequency response data for each plant. A total of 300 data points were used for soft clustering.



\begin{figure}[h!]
    \centering
    \includegraphics[width = 0.5\textwidth]{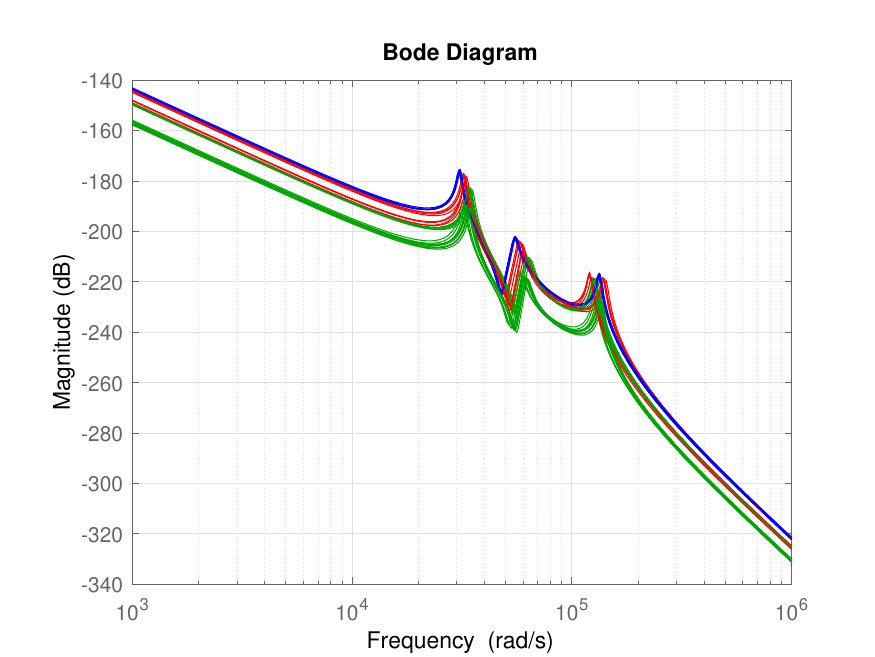}
    \caption{Clustering results for $30$ VCMs using k-medoids algorithm with $H_2$ distance respectively.}
    \label{fig:nyquistallplants}
\end{figure}


\section{Conclusion}\label{sec:conclusion}
In this paper, we presented a novel way of clustering transfer functions and frequency response data of LTI plants by defining distance metrics in terms of $H_\infty$ and $H_2$ norms. The distance metrics were used to cluster the systems using k-means and k-medoids algorithm. We also presented a novel way of soft clustering transfer functions and frequency response data utilizing Gaussian Mixture Models for a class of systems which are common in mechatronic systems like HDDs by extracting features from the frequency response data and applying using these features to cluster the frequency responses. The techniques have been demonstrated on various plants used in HDDs and the results are presented.



\end{document}